\documentclass[10pt]{ieeeconf}  

\usepackage{graphicx} 
\usepackage[letterpaper, margin=1in]{geometry}
\usepackage{hyperref}
\usepackage{xcolor}
\usepackage[english]{babel}
\usepackage{graphicx}
\usepackage{subcaption}
\usepackage{amsmath,booktabs}
\usepackage{pdflscape} 
\usepackage{makecell}
\usepackage{tabularx}
\usepackage[T1]{fontenc}
\usepackage{rotating,siunitx} 
\usepackage{newtxmath}
\newcommand{\rev}[1]{{\color{black} #1}} 

\title{Resilient Islanded Microgrid Battery Energy Management Considering the Life of Battery}
\title{Strategies for Resilience and Battery Life Extension in the Face of Communication Losses for Isolated Microgrids}
\author{Mohammad Hossein Nejati Amiri$^{\dagger}$, Fawaz Annaz$^{\dagger}$, Mario De Oliveira$^{\dagger}$, and Florimond Gueniat$^{\dagger}$
\thanks{$^{\dagger}$All authors are with the College of Engineering, Birmingham City University, Birmingham, UK
        {\tt\small mohammadhossein.nejatiamiri@mail.bcu.ac.uk\\ fawaz.annaz@bcu.ac.uk\\mario.deoliveira@bcu.ac.uk\\florimond.gueniat@bcu.ac.uk}}
}

\date{\textit{\today}}      

\begin{document}
\maketitle
\begin{abstract} \label{sec-abstract}
This study addresses the challenges of energy deficiencies and high impact low probability (HILP) events in modern electrical grids by developing resilient microgrid energy management strategies.
It introduces a sliding Model Predictive Control (MPC) methodology integrated with Battery Energy Storage Systems (BESS), emphasizing extending battery life and prioritizing critical loads during HILP events. 
This approach focuses on extending the sustainability of battery operation by linearizing the battery lifecycle within the optimization framework.
Furthermore, this research proposed a straightforward method to mitigate communication disruptions during HILP events, thereby ensuring operational integrity. 
This focused approach enhances isolated microgrid resilience and sustainability, offering a strategic response to contemporary environmental challenges.

\end{abstract}
\begin{keywords} 
Microgrid, Resiliency, Battery Energy Management, Model Predictive Control, Renewable Energy Resources 
\end{keywords}

\section{Introduction} \label{sec-Introduction}

Contemporary power systems face two serious challenges: energy shortages and the escalation of HILP events like natural disasters and cyber-attacks.
Smart grids (SGs) have been introduced as a solution to tackle these challenges sustainably and efficiently \cite{Resilience_Energy_deficiencies}. 

The Advanced Metering Infrastructure (AMI), a fundamental component of SGs, facilitates innovative concepts such as demand response, prosumers, and microgrids \cite{Nejati}.
This bidirectional communication system revolutionizes energy exchange dynamics by efficiently utilizing renewable energy sources (RES). Among these concepts, microgrids particularly stand out for their ability to efficiently harness RES, offering a clean solution to energy deficiencies.
This is particularly crucial for remote areas where extending power lines is impractical, and RES can offer a viable alternative.
Microgrids are capable of operating in both connected and isolated modes \cite{Isolated_MG}.
The isolated mode is considered here to serve remote areas.

The inherent intermittency of RES poses a significant challenge to the stability of microgrid systems. 
BESS plays a critical role in managing the energy fluctuations within isolated microgrids by storing excess energy during periods of surplus generation and supplying it during increased demand \cite{BESS_Importance}.
Given their high cost, extending the lifespan of BESS is paramount for their long-term viability within the system \cite{BESS_Cost}. 
Hence, this paper proposes a method of directly incorporating the linearization of Depth of Discharge (DOD) to Battery Life Cycle (BLC) curves into the MPC optimization problem to maximize BESS lifespan.

Furthermore, in the face of HILP events, prioritizing the supply of critical loads over regular loads becomes essential \cite{Resilient_essential}.
This study integrates such prioritization within the MPC model. 
Additionally, during HILP events, SGs face the risk of communication loss with loads. 
This paper also considers and mitigates the impacts of potential communication failures by predicting load demand during critical hours.

Recent advancements in Model Predictive Control (MPC) for isolated microgrid energy management, such as those presented in \cite{shehzad2021}, have explored optimizing battery life through minimizing switching costs. However, these efforts do not directly incorporate battery longevity in their objective functions, leaving uncertain impacts on actual battery life. Moreover, while studies like \cite{MPC_Resilience_1}, \cite{MPC_Resilience_2}, \cite{MPC_Resilience_3}, and \cite{MPC_Resilience_4} advance distributed control, cooperative energy management, load prioritization, and outage management to enhance resilience and fault tolerance, they largely neglect the critical issues of communication system resilience and battery sustainability. This oversight in the literature highlights a significant gap in effectively addressing communication failures and ensuring battery longevity within MPC models for achieving comprehensive microgrid resilience.

The proposed methodology adopts Python Optimization Modeling Objects (PYOMO) for our modeling framework, a choice motivated by PYOMO's robust modeling capabilities and its versatility with various solvers. 
By making this strategic decision, this paper leverages Gurobi's optimization capabilities, which is renowned for its performance and efficiency \cite{Pyomo,gurobi}.

This study introduces several advancements in the optimal operation of isolated microgrids.
Optimal operation of isolated microgrids is carried out using a sliding window mechanism MPC model, which significantly improves decision-making accuracy by incorporating future data into hourly analyses, thus mitigating the impact of initial conditions.
Moreover, by linearizing the BESS lifecycle within the optimization's objective function, it optimizes the computational burden while focusing on extending BESS lifespan as a key step toward sustainable energy management. 
The study broadens its approach to energy distribution by addressing both standard and resilient operations, employing linearization techniques to reduce power imbalances and ensure a resilient system. 
Additionally, a prioritization scheme is introduced to distinguish between essential and regular loads, particularly during HILP events, ensuring that critical services maintain functionality. 
Load prediction mechanisms are integrated to counter potential communication disruptions during such events, enhancing the model's realism and resilience.
This approach ensures the system's robustness in facing unforeseen disruptions.

The structure of this paper is as follows. 
Section \ref{sec-MG_model} introduces the conceptual framework of the research, and will focus on conventional and resilience-enhanced control mechanisms for BESS within a microgrid architecture.
Section \ref{sec-SIM_result} examines simulation findings, evaluating the performance of the upgraded MPC model and the effects of innovations such as load prediction and priority.
Finally, section \ref{sec-Conclusion} will provide analysis of the study and its impact on enhancing the sustainability, resilience, and efficiency of isolated microgrids. 
It will also present guidance for future research directions. 

\section{Micro-grid Model} \label{sec-MG_model}
\begin{figure} 
    \centering
    \includegraphics[width=0.48\textwidth]{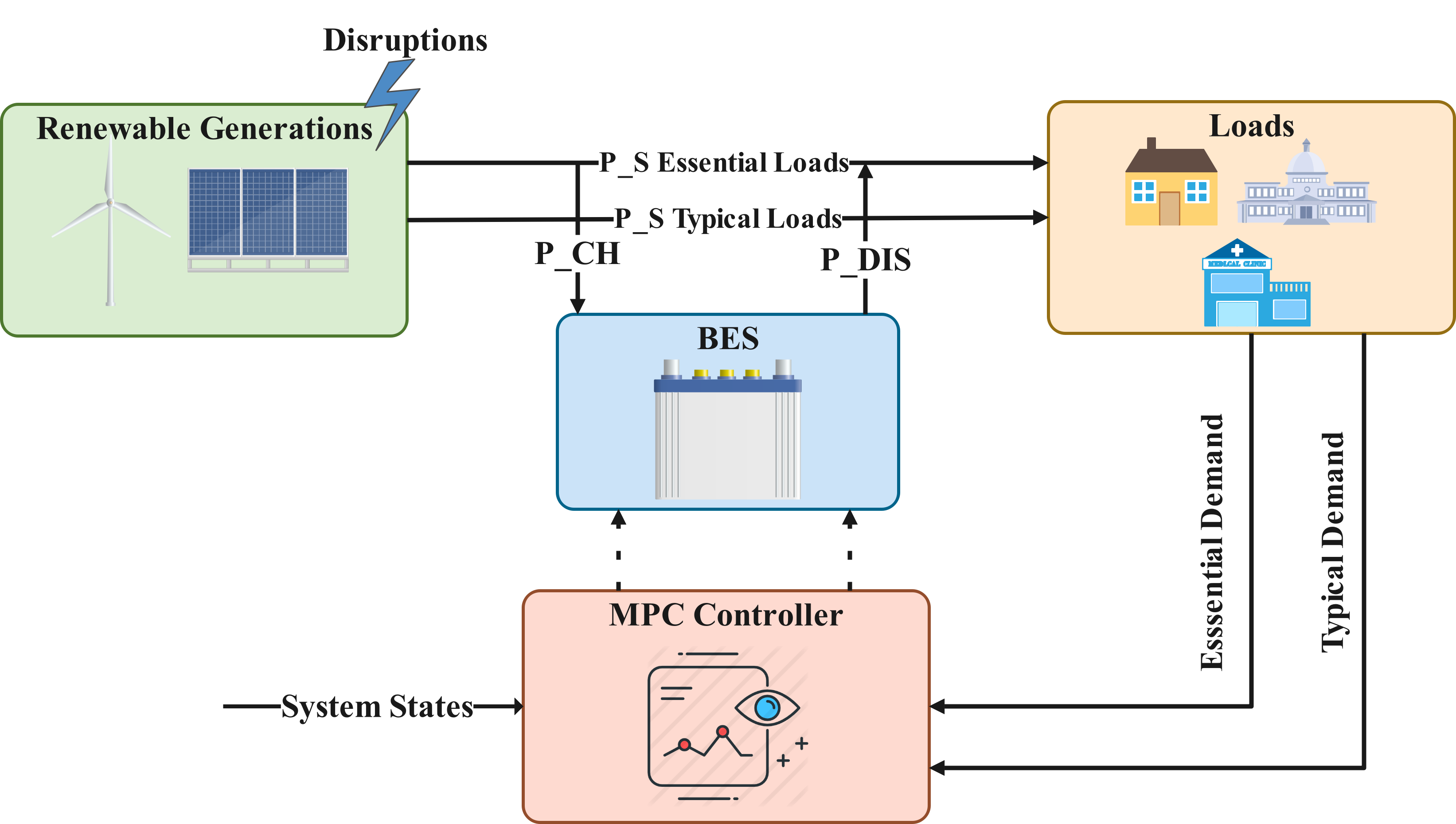} 
    \caption{Proposed Microgrid Energy Management Architecture.}
    \label{fig:model}
\end{figure}

This section discusses the components of the model and the mathematical formulation behind it.
As illustrated in Figure \ref{fig:model}, the architecture of the microgrid encompasses RESs such as wind and solar generators, alongside BESS facilities. 
The design incorporates dual power lines dedicated to supplying both essential and regular loads independently. 
Furthermore, this figure highlights the role of the MPC controller, which is an essential component responsible for regulating the battery's charging and discharging processes.
\rev{In this model, the MPC is assumed to be on the supplier side, managing the renewable generation and battery storage.}

\begin{figure}
    \centering
    \begin{subfigure}{.48\textwidth}
        \centering
        \includegraphics[width=1\textwidth]{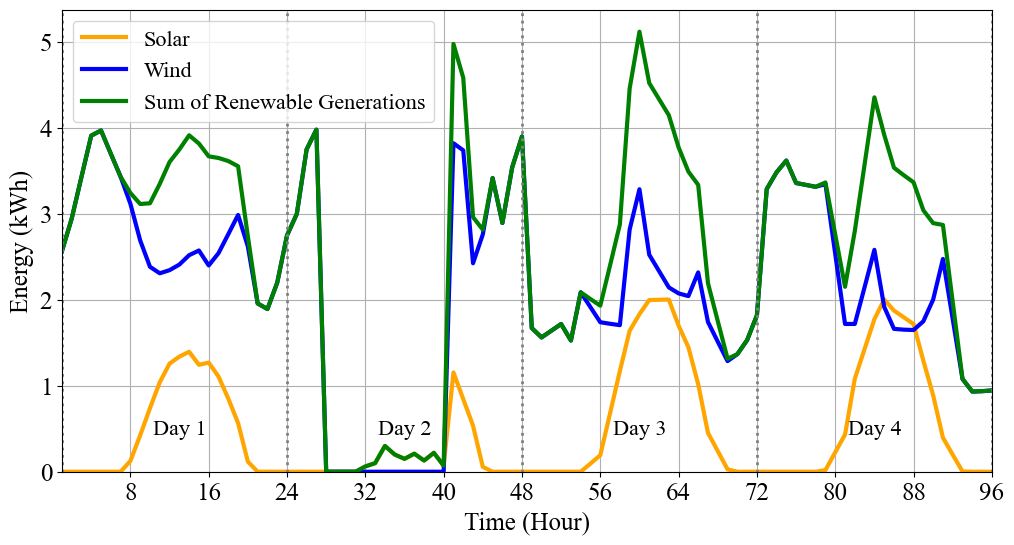}
        \caption{}
        \label{fig: RE_GEN}
    \end{subfigure}
    
    \begin{subfigure}{.48\textwidth}
        \centering
        \includegraphics[width=1\textwidth]{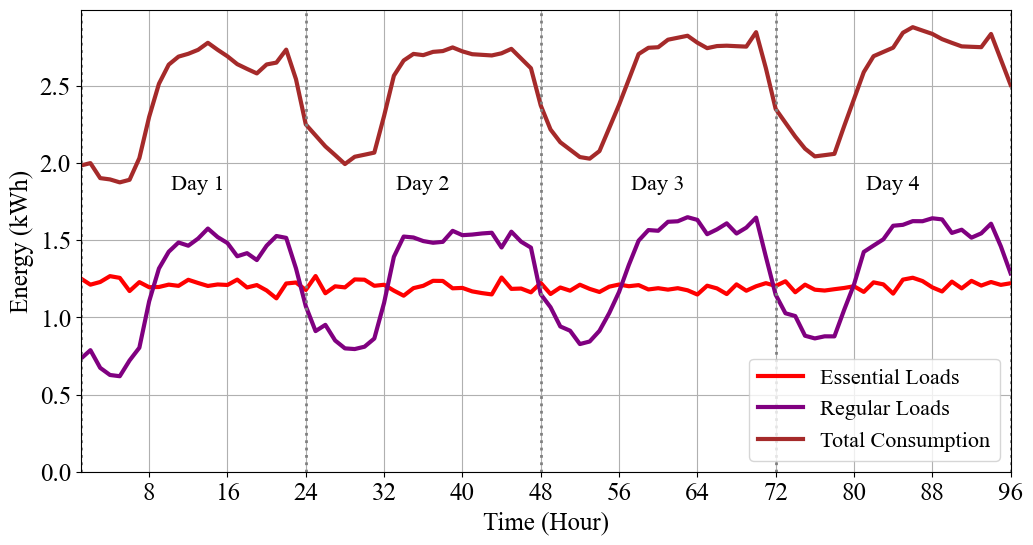}
        \caption{}
        \label{fig: Load_CONS}
    \end{subfigure}

    \begin{subfigure}{.48\textwidth}
        \centering
        \includegraphics[width=1\textwidth]{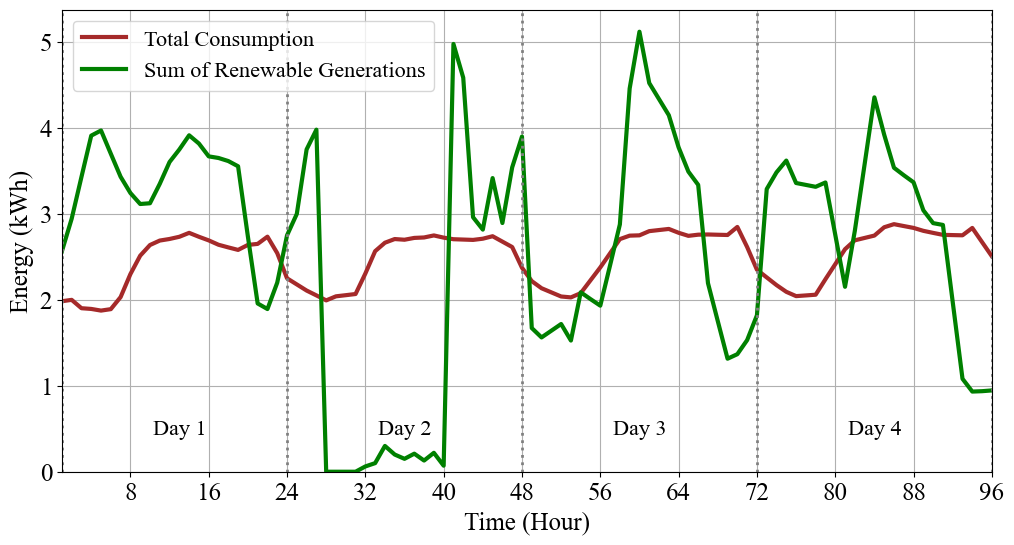}
        \caption{}
        \label{fig: GEN_CONS}
    \end{subfigure}
    
    \caption{a) Wind, Solar, and Total Power Generation. b) Essential Loads, Regular Loads, and Total Load Consumption. c) Four-Day Total Consumption and Renewable Generation.}
    \label{fig: Consumption_Generation}
\end{figure}

Taking into consideration the generation sources, Figure \ref{fig: RE_GEN} shows a typical output from solar and wind sources over four days and their summation. 
This timeframe exceeds the three days considered by the sliding MPC to accommodate its operational logic, which involves anticipating the next 24 hours for each battery management decision, and then sliding the assessment window by 24 hours following each decision.
Thus, for a three-day analysis, data for four days is essential.
Notably, in the event of a HILP scenario like a wind storm, wind generation typically peaks just before and after the event. 
In fact, during a storm, wind turbines may cease operation due to safety mechanisms, leading to a drop in wind generation to zero, while some solar generation might still occur.
In this case, in the second day, between hours 27 to 39, the system was hit by a HILP event.
Figure \ref{fig: Load_CONS} illustrates the consumption patterns of essential and regular loads, along with their combined total. 
The essential load demonstrates a stable, nearly flat profile, fluctuating within a narrow range, in contrast to the regular load, which varies throughout the day. 
This analysis pertains to a small-scale system, predominantly characterized by residential or commercial usage.
Figure \ref{fig: GEN_CONS} presents the overall power demand and supply.

\begin{figure}
    \centering
    \includegraphics[width=0.48\textwidth]{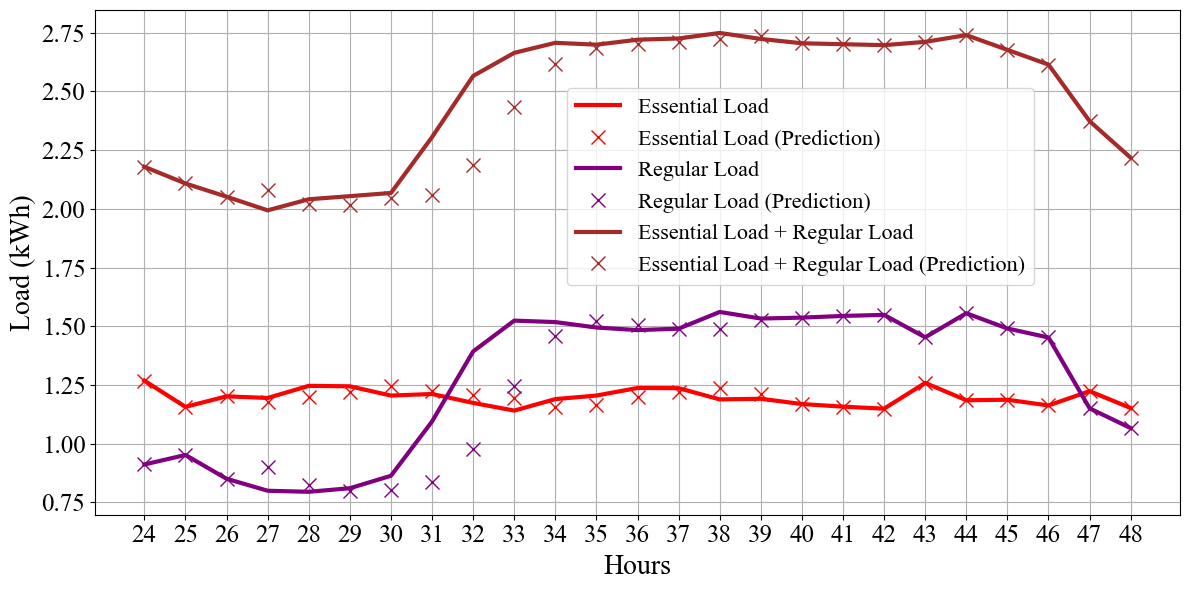}
    \caption{Essential and regular loads (straight lines) and load predictions (markers) during HILP event.}
    \label{fig:Prediction}
\end{figure}

Another significant contribution of this article lies in addressing the potential loss of the communication system, or the cyber layer, during the HILP event. 
Specifically, the disruption begins at hour 27, leading to the assumption that data on load consumption from the load side becomes unavailable. 
A simple yet effective strategy that uses the average of the last two hours to estimate the load demand was adopted. 
However, upon concluding hour 27 and moving to decide on hour 28, the exact consumption for hour 27 becomes known, allowing for the use of precise data from hour 27 combined with hour 26 data to forecast the load demand for hour 28.
Employing this strategy results in a Root Mean Square Error (RMSE) of 0.035 for essential loads, 0.162 for regular loads, and 0.145 for the total load.
The advantage of this method is the minimal error observed in predicting essential loads, thus enhancing the robustness of the optimization process, even in scenarios of cyber layer disruption. 
This improved accuracy is attributed to the typically stable consumption pattern of essential loads, characterized by minimal fluctuations.
The prediction for the second day is demonstrated in Figure \ref{fig:Prediction}.

\begin{figure}
    \centering
    \includegraphics[width=0.48\textwidth]{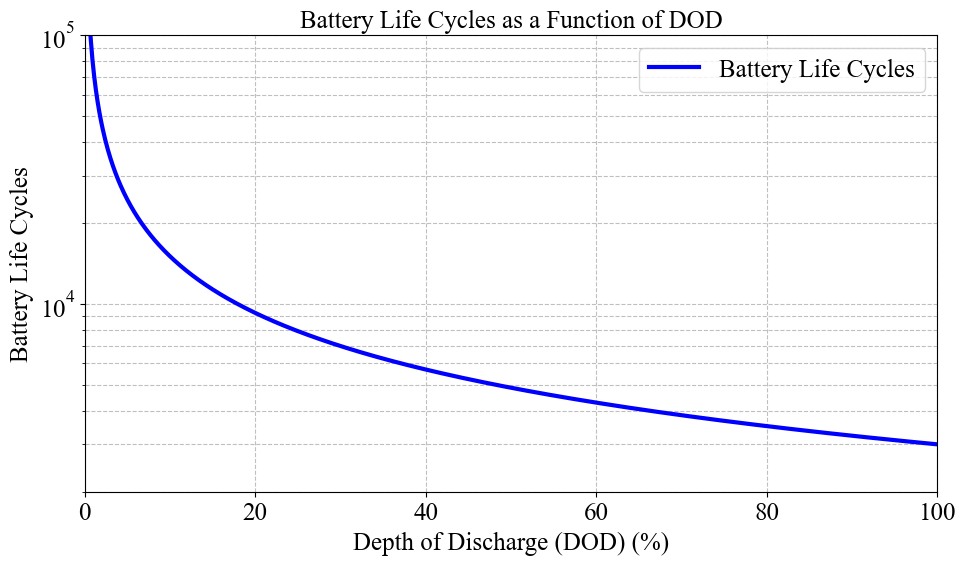}
    \caption{Battery Life Cycle as a Function of DOD.}
    \label{fig: BLC}
\end{figure}

This work evaluates the effectiveness of a conventional BESS control strategy against an integrated BESS resilience management approach.
Including an examination of the battery's life cycle within the resilient MPC model represents a significant advancement contributed by this research. 
The model, based on Mixed Integer Linear Programming (MILP), simplifies the curve of Depth of Discharge (DOD) related to battery lifecycle into eight linear sections. 
This adaptation facilitates its direct application in the MPC optimization problem, as shown in Figure \ref{fig: BLC}.
\rev{By manipulating several mathematical constraints, we force the MILP problem to choose from these eight segments based on the DOD variable. Consequently, the Battery Life Cycle (BLC) is calculated as part of the objective function, ensuring its integration into the optimization model while keeping the problem linear}

Now, a general mathematical optimization framework pertinent to this study will be introduced.
Model parameters are detailed in Table \ref{tab:model_parameters}.
\begin{table}[ht]
\centering
\caption{Resilient Energy Management of Microgrid Optimization Parameters}
\label{tab:model_parameters}
\begin{tabular}{lp{3.5cm}}
\hline
\textbf{Parameter} & \textbf{Initialization} \\ \hline
$P_{\text{load\_1}}$ & Essential Load Data \\
$P_{\text{load\_2}}$ & Regular Load Data \\
$P_{\text{RE\_GEN}}$ & Renewable Generation Data \\
$\eta_{\text{ch}}$ & 0.90 \\
$\eta_{\text{dis}}$ & 0.95 \\
$SOC_{\text{init}}$ & 0.5\\
$SOC_{\text{min}}$ & 0.2 \\
$SOC_{\text{max}}$ & 0.9 \\
$\omega_{\text{bat}}$ & [0,1] \\
$w_{t}$ & [0,1] \\
$w_{R}$ & [0,1] \\
$w_{BLC}$ & [0,1] \\
$Big\_M$ & 100 \\
$C_{\text{bat}}$ & 125 \\
$C_{\text{ch\_dis}}$ & 0.055 \\
$C_{\text{no\_ch\_dis}}$ & 0.055 \\
$C_{\text{idle}}$ & 0.0275 \\
$P_{bat_{\text{max}}}$ & 4 \\
\hline
\end{tabular}
\end{table}

Within the optimization framework, three primary sets are identified.
The set $K$ corresponds to time slots, encompassing a complete 24-hour cycle. 
The set $A$ details the operational states of the battery, such as charging (CH), discharging (DIS), and idle (IDLE) modes.
The operational state of the battery at any time period $k$ is governed by the constraint that exactly one of the states (charging, discharging, or idle) must be active.

This work builds upon the optimization framework outlined in \cite{shehzad2021}, incorporating formulas for the State of Charge (SOC) of BESS, various BESS operational modes, and BESS switching costs into the objective function $J_{\text{bat}}(k)$, alongside the total power imbalance cost function $J_t(k)$. 
The battery life cycle is calculated after optimization as a sub-product in \cite{shehzad2021}, whereas in this study, it is linearized and directly incorporated into the objective function.
Additionally, this study enriches the model by integrating resiliency and battery lifecycle considerations into the cost function, expanding the framework capacity for sustainable and resilient energy management.

The SOC rule, grounded in MPC, ensures that SOC is updated based on the net effect of charging and discharging activities. 
The MPC framework incorporates a sliding window approach to improve the accuracy of SOC forecasting.
This strategy involves utilizing data for a future 24-hour period at each hourly increment to update the SOC, ensuring decisions are informed by a comprehensive predictive outlook. 
Specifically, if the model is at hour 1, it employs data for the next 24 hours to determine the SOC for hour 2. 
As the model advances to hour 2, it then shifts the window to incorporate the subsequent 24-hour dataset for calculating the SOC for hour 3, and this process continues. 
This sliding window mechanism facilitates an ongoing forward-looking analysis, enabling the model to make well-informed predictions and optimizations.
Although this method provides data extending to day 4, the optimization results are confined to 3 days.
This modelling technique highlights the anticipatory nature of MPC by optimizing future states based on current decisions and available forecasts, thus significantly influencing the system’s energy management strategy through an accurate reflection of the battery's operational dynamics over a predictive horizon. 
Additionally, this approach mitigates the impact of the choice of initial SOC, making the optimization process less sensitive to initial conditions and more resilient to variations in starting SOC levels.

A key element of the proposed optimization framework is the management of the battery's life cycle, which relies on a clear understanding of how batteries degrade over time. 
This is achieved through a structured piecewise linear model that encapsulates the relationship between the DOD and the BLC.

\rev{The total cost associated with the battery life cycle for each period, $k$, is represented by $J_{BLC}(k)$. 
This total cost is a summation of individual life cycle costs across all segments, $b$, within the set $B$, indicative of the battery's degradation stages. 
The set $B$ depicts the eight distinct sections of the battery life cycle curve, with each section represented as a piecewise linear segment. 
Mathematically, this relationship is expressed as:

\begin{equation}
J_{BLC}(k) = \sum_{b \in B} BLC(b, k)
\end{equation}

where $BLC(b, k)$ denotes the cost attributed to the battery life cycle for segment $b$ at time $k$. }

The framework's optimization includes a cost function for power resiliency, $J_R(k)$, which underscores the importance of prioritizing essential loads over regular loads. 
This prioritization is mathematically articulated in the resiliency cost function at time $k$ as:
\begin{subequations}
    \begin{align}
        J_R(k) = 1 - \left(\frac{\text{Total\_weighted\_loss}(k)}{\text{Total\_weighted\_load}(k)}\right). \label{eq:resiliency_cost_function} \\
        \text{Total\_weighted\_loss}(k) = \nonumber\\ \omega_{\text{essential}} \times P_{\text{essential\_imbalance}}(k) \nonumber \\
        + \omega_{\text{regular}} \times P_{\text{regular\_imbalance}}(k). \label{eq:total_weighted_loss}
    \end{align}
\end{subequations}
Here
$P_{\text{essential\_imbalance}}(k)$ indicates the power imbalance for essential loads at time $k$, while
$P_{\text{regular\_imbalance}}(k)$ signifies the power imbalance for regular loads at time $k$.
$\omega_\text{essential}$ and $\omega_\text{regular}$ are coefficients reflecting the relative importance or cost of power imbalances in essential and regular loads, respectively.
Assigning $\omega_\text{essential}$ a higher value than $\omega_\text{regular}$ quantifies the prioritization of essential loads, ensuring that disruptions affecting critical operations carry a greater cost. 
This strategic allocation of weights effectively guides the system toward maintaining service continuity for essential functions.

The cost function associated with the total power imbalance, denoted as $J_t(k)$, is crucial for evaluating the efficacy of power distribution within the system.
At any given time $k$, this cost is quantified by the squared difference between the total power supply and the sum of loads from both essential and regular categories.
The square of the difference ensures that any deviation from the balance, either surplus or deficit, is penalized, thereby promoting a closer alignment of supply with demand across all types of loads.

The objective function of the optimization model aims to minimize the overall cost, taking into account various components related to battery operation, BLC, power imbalance, and power resiliency. 
This function is formulated as follows:
\begin{equation}
\begin{aligned}
\min \sum_{k \in K} & \big( \omega_{\text{bat}} \cdot J_{\text{bat}}(k)  - \omega_{\text{BLC}} \cdot J_{\text{BLC}}(k) \\
& + \omega_{\text{t}} \cdot J_t(k) - \omega_{\text{R}} \cdot J_R(k) \big),
\end{aligned}    
\end{equation}
where
$\omega_{\text{bat}}$, $\omega_{\text{BLC}}$, $\omega_{\text{T}}$, and $\omega_{\text{R}}$: are weighting factors for the respective costs, reflecting their relative importance in the objective function.
While the first two terms of the objective function control the battery switching and life cycle, the latter two elements deal with power management.
This objective function integrates the various cost components, ensuring a balanced operational efficiency, battery sustainability, power balance, and resiliency within the energy management system.

\section{Simulation Results and Discussions} \label{sec-SIM_result}
In this section, two distinct scenarios will presented. The first incorporates all objective functions with precise load data, and the second addresses the loss of the communication system during a HILP event.

\begin{figure}
    \centering
    \begin{subfigure}{.48\textwidth}
        \centering
        \includegraphics[width=1\textwidth]{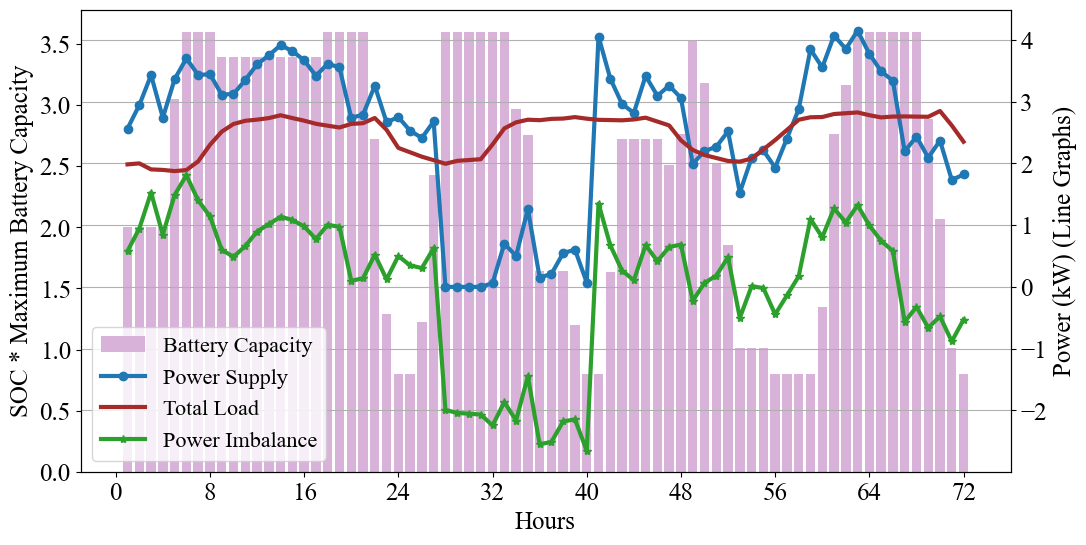}
        \caption{}
        \label{fig: Total_imbalance}
    \end{subfigure}
    
    \begin{subfigure}{.48\textwidth}
        \centering
        \includegraphics[width=1\textwidth]{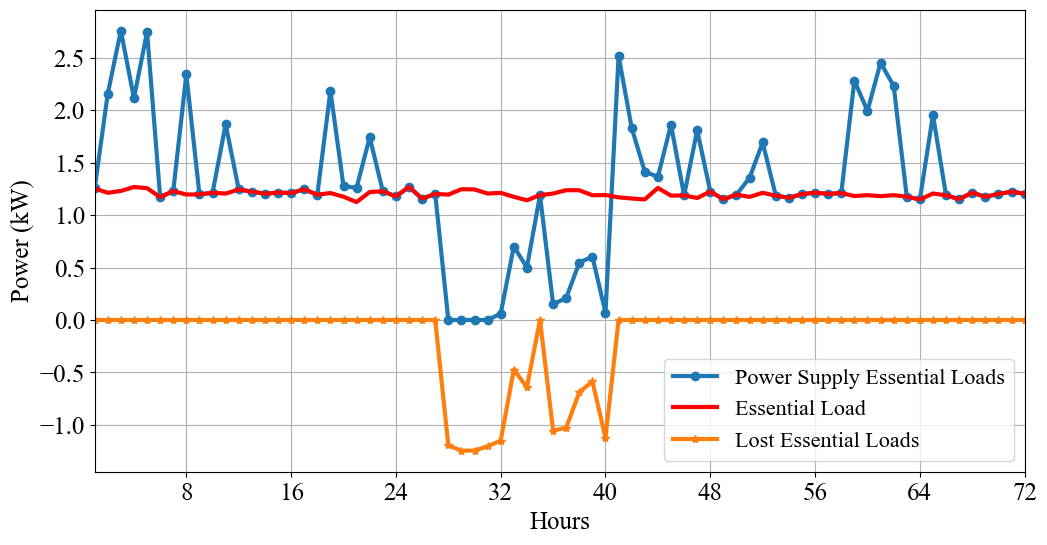}
        \caption{}
        \label{fig: Essential_imbalance}
    \end{subfigure}
    \begin{subfigure}{.48\textwidth}
        \centering
        \includegraphics[width=1\textwidth]{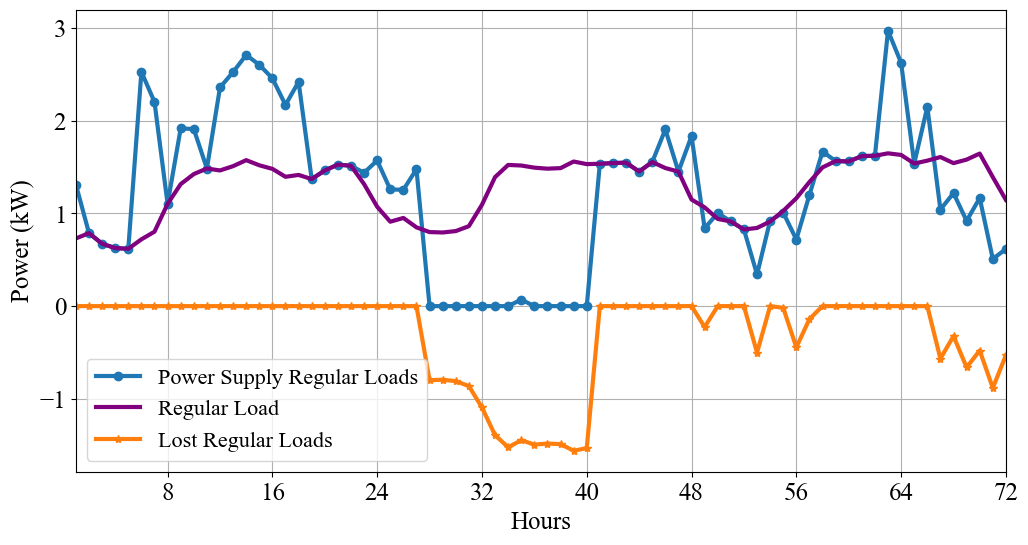}
       \caption{}
        \label{fig: Regular_imbalance}
    \end{subfigure}

    \caption{
    a) Total Power Imbalances and State of Charge. 
    b) Power Imbalances in Essential Loads.
    c) Power Imbalances in Regular Loads.}
    \label{fig: Imbalance for All Objectives}
\end{figure}

Upon comparing various scenarios that involved adjusting the weighting of elements within the objective function, the scenario in which all elements were considered emerged as the most balanced. 
This scenario achieved equilibrium, characterized by reasonable total switches, expected lifespan, resilience index, and system dynamics.
It resulted in a total of 22 switches between different operation modes of the BESS, with the battery discharging 19 times.
Essential load losses amounted to 11.629 kW, and regular load losses to 21.061 kW, culminating in a total loss of 32.691 kW.
The resilience index for this scenario stood at 0.845, and the expected lifespan of the battery was estimated at 29.29 years, as shown in Figure \ref{fig: Imbalance for All Objectives}.
Identified as the optimal approach for microgrid energy management, the second scenario takes into account the potential loss of the cyber layer during HILP events.
The second scenario considers a loss of communication with loads during an HILP event. 
The straightforward forecasting method proposed in section \ref{sec-MG_model} for load prediction during these hours exhibits commendable accuracy, with an RMSE for essential loads around 3.5\%, indicating nearly identical resilience index outcomes with a minor difference of 0.001 compared to scenarios with precise load knowledge. 
As outlined in Equation \ref{eq:total_weighted_loss}, the weighting factor for essential loads is higher. 
Due to improved accuracy in predicting essential load consumption, the resilience index (RI) remains consistent.
However, there is a greater loss in regular loads, attributed to lower precision in their prediction.
The outcomes for the second scenario closely align with those of the first scenario, revealing a one-year reduction in battery life when load data is not precisely known.\\
In the second scenario, losses in essential loads were recorded at 11.559 kW, while increased in regular loads at 21.621 kW, resulting in a resilience index of 0.844. 
The expected battery lifespan is estimated at 28.372 years, reflecting a one-year decrease compared to the first scenario. 
Additionally, this scenario observed 15 switches in operation modes of the BESS, with the battery discharging 23 times.

Delving into the specifics of the first scenario, which emerges as the optimum choice, Figure \ref{fig: Total_imbalance} displays the battery's SOC over three days.
Observations reveal that prior to hour 28, when the HILP event impacts the system, the battery charges itself to reach maximum capacity.
It then endeavors to sustain this charge level to minimize essential load loss during the latter hours of the HILP event. 
In this scenario, it becomes evident that the management strategy is mindful of battery longevity, opting to keep the battery in IDLE mode until its use is necessary.
It is noteworthy that the battery's operational capacity remains constrained within a range from 0.2 kW to 3.6 kW due to the SOC limitations imposed on the battery system.


Figure \ref{fig: Imbalance for All Objectives} displays the power supply, total demand, and total imbalances as shown in \ref{fig: Total_imbalance}. 
It is important to note that positive imbalances occasionally occur due to the limited capacity of the BESS. 
Furthermore, Figure \ref{fig: Essential_imbalance} illustrates the optimal management's efforts to minimize losses as much as possible. 
On the other hand, Figure \ref{fig: Regular_imbalance} indicates that power management prioritizes essential loads over regular ones, resulting in less attention to the latter.

\begin{figure}
    \centering
    \includegraphics[width=0.48\textwidth]{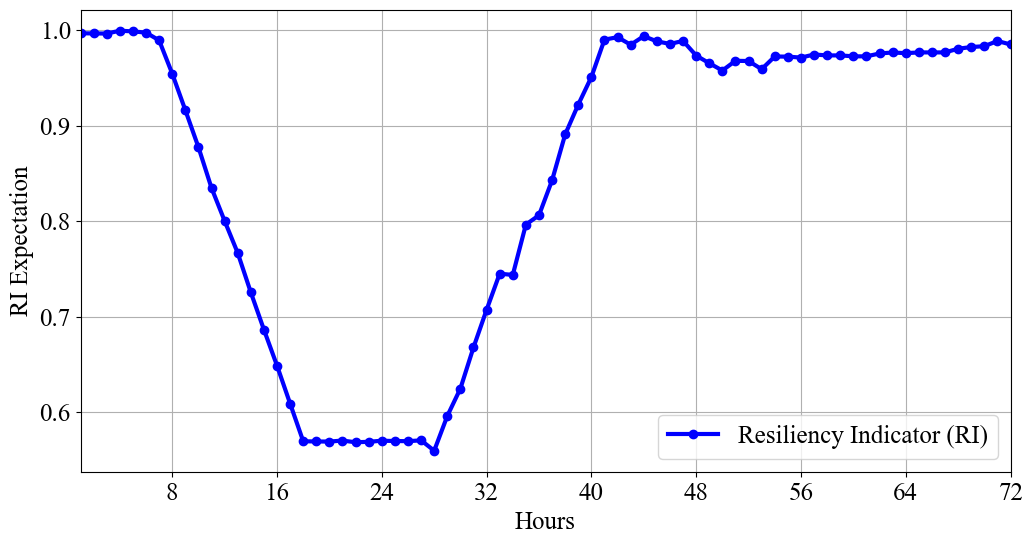} 
    \caption{Resilience Index Expectation in Sliding MPC.}
    \label{fig:Expected_RI}
\end{figure}

The expected RI curve is utilized to evaluate the impact of the sliding MPC approach presented, as illustrated in Figure \ref{fig:Expected_RI}. 
This is evident when an HILP event occurs at hour 3 on day 2, persisting for 13 hours within the system. 
Consequently, the expected RI diminishes, reaching its lowest point during the MPC decision-making process at hour 17. 
This decline continues for several hours, attributed to the MPC's forward-looking capability. 
Upon anticipating a return to normalcy, the MPC forecasts a subsequent upturn in the RI trend.

\section{Conclusion} \label{sec-Conclusion}
This study highlights the effectiveness of a sliding MPC approach, utilizing PYOMO and Gurobi for optimizing microgrid energy management, to enhance sustainability and resilience.
By incorporating future data for decision-making, the model significantly improves prediction accuracy and operational efficiency, particularly by extending BESS longevity and prioritizing essential loads during HILP events.
Results demonstrate improvements in resilience and battery life, contributing to advancements in isolated microgrid optimization. 

Future research will focus on integrating flexibility resources like demand response, addressing uncertainties in renewable generation and load demand with advanced optimization techniques, and developing precise load prediction models. \rev{Additionally, we will compare our results with multi-objective optimization approaches, consider real-world case studies including real communication protocols and hardware-in-the-loop setups.}
These initiatives aim to elevate the resilience, sustainability, and efficiency of microgrid management.
\bibliographystyle{ieeetr}
\bibliography{root}

\end{document}